# TDoA Based Positioning using Ultrasound Signals and Wireless Nodes

Alessio De Angelis, Antonio Moschitta, Antonella Comuniello
Dept. of Engineering
University of Perugia
Perugia, Italy
{alessio.deangelis,antonio.moschitta}@unipg.it

*Abstract*—In this paper, a positioning technique based on Time Difference of Arrival (TDoA) measurements is analyzed. The proposed approach is designed to consent range and position estimation, using ultrasound transmissions of a stream of chirp pulses, received by a set of wireless nodes. A potential source of inaccuracy introduced by lack of synchronization between transmitting node and receiving nodes is identified and characterized. An algorithm to identify and correct such inaccuracies is presented.

## I. INTRODUCTION

Ultrasound transmissions are a well-known and deeply studied technique, with applications ranging from biomedical scanning to industrial and automotive applications [1]-[3]. Positioning techniques have been studied as well in the literature, because ultrasound transmissions consent accurate short range distance measurement and positioning, using low-cost and low-power hardware [4]-[12]. Recently, an ultrasound system for indoor positioning has been proposed in [13]. Such solution is based on handheld consumer devices using the Android platform and is capable of real-time operation with decimeter-order accuracy. Furthermore, in [14], a positioning system that allows for measuring range and bearing has been proposed, achieving an accuracy better than 10 cm in multipath environments.

Typical ultrasound positioning systems are based on time domain measurements, such as Time of Flight (ToF) and Time Difference of Arrival (TDoA) between the mobile node and a set of know position anchors. These measurements can either directly feed a positioning algorithm or they can be converted into range estimations using knowledge of speed of sound, prior feeding a lateration algorithm.

It is worth noticing that, while several ultrasound positioning systems have been proposed in the literature, most of them use some synchronization scheme. In some systems wired connection between anchors and the mobile node are used, while in some other cases wireless nodes are synchronized using industrial oriented radio protocols, such as ZigBee or Radio Frequency Identification (RFID) [15][16][17]. Wireless implementations are often a desirable solution, since wireless nodes can be easily deployed and flexibly relocated when installing and operating a positioning system. Moreover, recently several low power chips capable of radio communication were proposed on the market implementing the 4.0 Bluetooth low power protocol, also known as Bluetooth Low Energy (BLE) [18]. BLE is a good candidate for wireless implementations of ultrasound positioning systems. Not only BLE solutions are low cost, but they are usually implemented in hand held devices such as smartphones. Hence, using BLE as communication infrastructure to operate a distributed positioning system based on ultrasound techniques may consent to easily implement and support user-oriented Location Based Services. The main drawback of BLE is in its very same user-oriented nature. In fact, protocols like BLE are transparent to the user, and cannot easily be programmed and configured. In particular, BLE networks use adaptive frequency hopping Time Division Multiple Access (TDMA), where the hopping sequence is often hardware coded. Thus, the frequency hopping random latencies may not be compatible with the synchronization accuracy required by an ultrasound ToF positioning system. This issue may be overcome by realizing a TDoA system, that works in absence of synchronization.

Consequently, in this paper an ultrasound based TDoA positioning architecture is investigated, assuming that the mobile node acts as an active beacon, while fixed anchors act as listeners. The performance of the proposed TDoA approach is analyzed. It is shown that TDoA may be prone to uncertainty because of ambiguities in measuring time delays, that can be identified and removed using a proper algorithm.

## II. ARCHITECTURE OF THE ANALYZED SYSTEM

### A. System architecture and signaling

The proposed approach is summarized by Fig. 1, showing a mobile node acting as active beacon, a set of wireless anchors, acting as listeners, and a Master node, acting as supervisor. All nodes feature a radio interface, assumed to be a BLE one. All nodes but the Master are equipped with an ultrasound transceiver. The anchors are assumed to be time-synchronized with each other, but no synchronization is assumed between the mobile node and the anchors. The anchor synchronization may be achieved using a wireless protocol, as illustrated in Fig. 1(a), or by wired connection, as in Fig. 1(b).

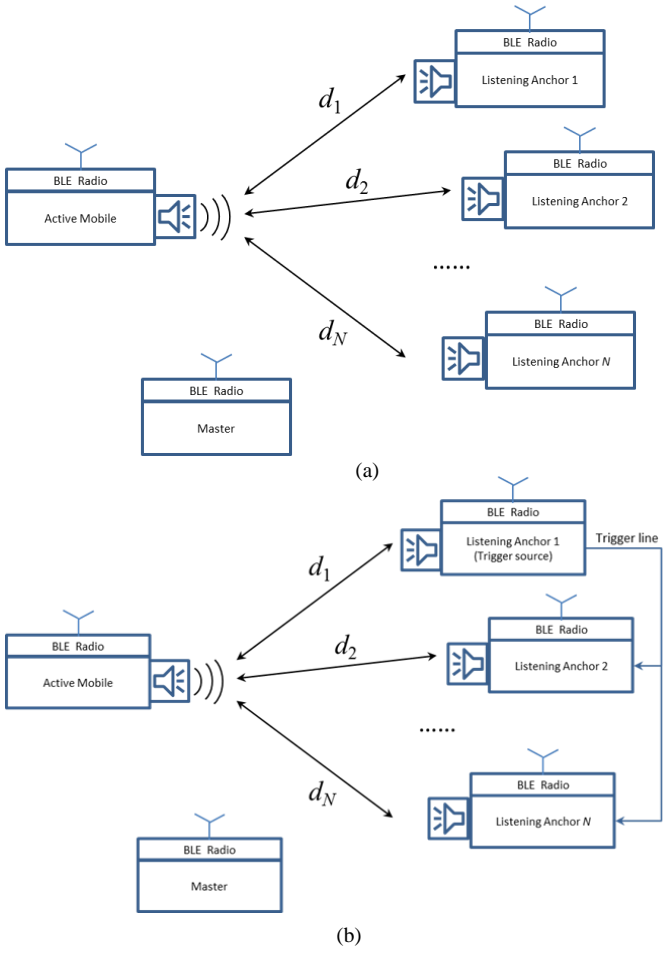

Fig. 1 – The considered positioning system. Two possible implementations are depicted: in (a) the anchor nodes are synchronized with each other using a wireless protocol. In (b), the anchor nodes are synchronized by means of a wired connection providing the trigger signal.

When a position is to be estimated, the Master node triggers both the mobile node and the anchors using the radio interface. The mobile node begins transmitting a repetitive sequence of ultrasound pulses, while the anchors start listening for incoming signals at their own ultrasound transceiver output. The number of pulses and the pulse duration are suitably selected to ensure that any latencies due to RF transmissions do not prevent the reception of ultrasound pulses. To this aim, it should be observed that Bluetooth systems operate with a hopping frequency of 1600 hops/s, corresponding to a time slot of 625μs. Since the speed of sound in air is about 343 m/s, the delay corresponding to a BLE time slot is equivalent to a ranging error of about 21 cm. As in typical BLE chips the slot allocation cannot easily be controlled, a pseudorandom latency of several time slots can be introduced, unsuitable for ranging and positioning systems based on ToF measurements.

As anticipated in the introduction, TDoA works in absence of synchronization between the mobile node and the anchors, assuming synchronization between the anchors. The anchors compare their instant of reception of the received ultrasound pulse, measuring the reciprocal delays. Then, the measured delays are used to infer the position of the mobile node, by solving the system of equations given by

$$v\tau_{ij} = R_i - R_j, \quad i, j = 1,...,N, \quad i \neq j. \quad (1)$$

where $N$ is the number of anchors, $v$ is the speed of sound in air, $\tau_{ij}$ is the TDoA between the $i$-th node and the $j$-th node, and $R_m$ is the distance between the $m$-th anchor and the mobile node, given by

$$R_m = \sqrt{(x_m - x)^2 + (y_m - y)^2 + (z_m - z)^2}, \quad (2)$$

where $(x_m, y_m, z_m)$ are the coordinates of the $m$-th beacon, while $(x,y,z)$ are the coordinates of the mobile node. In the considered system, each anchor acquires the output of its ultrasound microphone for a time window of duration $T_W$, chosen so as to guarantee that a transmitted ultrasound pulse can be collected by the various anchors. Each anchor can measure the time of arrival of the incoming signal using correlation techniques on a stored template of the transmitted ultrasound pulse. Throughout this paper, the mobile node is assumed, without loss of generality, to transmit a continuous train of linear chirp pulses, given by

$$s(t) = s_0\left(T_C \left\langle \frac{t}{T_C} \right\rangle \right)$$

$$s_0(t) = \begin{cases} A\sin(2\pi f(t)t), & f(t) = f_0 + \frac{f_1 - f_0}{2T_C}t, \quad 0 \leq t \leq T_C \\ 0, & \text{elsewhere} \end{cases} \quad (3)$$

where $f_0$ is the lowest frequency, $f_1$ is the highest frequency, $T_C$ is the duration of a single chirp pulse, and $\langle \cdot \rangle$ is the fractional part operator.

The choice of chirp signaling is due to the good correlation properties of such signals, and to the availability of interpolation techniques that consent to improve the time measurement resolution with respect to the sampling time of the adopted data acquisition system [11][12].

As described previously, the function of the Master node is to initiate remotely the measurement procedure. It may be noticed that its function might be integrated in one of the anchors or in the mobile node. Alternatively, a separate node for initiating the procedure remotely could be employed, as depicted in the diagram of Fig. 1. The choice of whether to integrate the function of the Master in one of the other nodes or in a separate remote node is application-dependent.

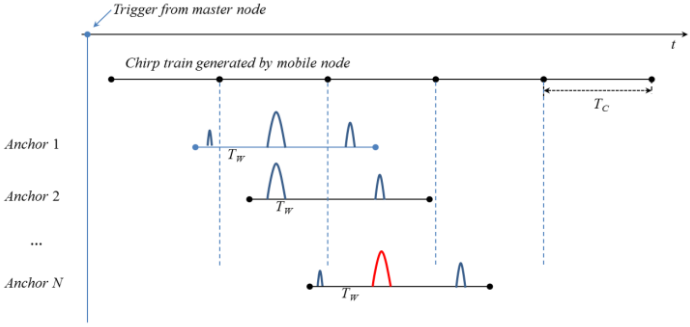

Fig. 2 – Time diagram of chirp detection in the considered system. The curves depict the correlation peaks originated by portions of consecutive chirp pulses entering the observation window of each anchor. The largest peak observed by the last node, highlighted in red, differs from the largest peak observed by the other nodes.

### B. A potential source of ambiguities

It is worth noting that the TDoA approach operates under the implicit assumption that all the receivers measure and compare the time of arrival of the same pulses. In order to properly detect a pulse, the $T_W$, the duration of the observation, must be at least equal to $T_C$. However, since in the considered wireless scenario the acquisition is not synchronized to the transmissions, in the typical case at least two consecutive replicas of the transmitted chirp will be partially acquired, as shown in Fig. 2. If the collected record described in Fig. 2 is processed by correlating it with a stored template, multiple correlation peaks may occur, one for each of the collected partial replicas. Similar problem occurs when computing the cyclical correlation between the template and the acquired signal.

When multiple receivers are involved, as expected for a positioning system, the TDoA will operate correctly only if of all of them select the correlation peak corresponding to the same pulse. Depending on the sources of latency, on the correlation algorithm and on the peak selection strategy, this may not occur. If each receiver evaluates the correlation sequence between the received signal and the stored template and selects the highest correlation peak among the available ones, depending on latencies and noise some receiver may select a peak belonging to a different pulse replica than the remaining anchors. This case is summarized by the last node in Fig. 2. This behavior was analyzed by meaningful simulations. For instance, Fig. 3, obtained in a planar scenario, shows the probability that TDoA measurements from two anchors, (represented by circlets) are affected by the considered ambiguity. The results were derived for a chirp duration of $T_C$=15 ms and assuming $T_W = T_C$. For each position in a 2-cm-pitch grid in an area of 4 m by 4 m, the simulation was repeated varying the latency between the master trigger and the start of the observation window from 0 to $T_C$ in 0.5 ms-steps. An error event is generated when the difference between the measured TDoA and the true one is larger than 30 us, corresponding to a 1-cm error.

Notice that the delay between the ultrasound receivers is not problematic in the considered system architecture, because the ultrasound receivers are assumed to be synchronized with each other, possibly using wireless synchronization protocols such as those analyzed in [19]-[21]. With respect to a fully synchronized solution, the proposed approach consents using a simple unsynchronized mobile node, relaxing its processing and power consumption requirements.

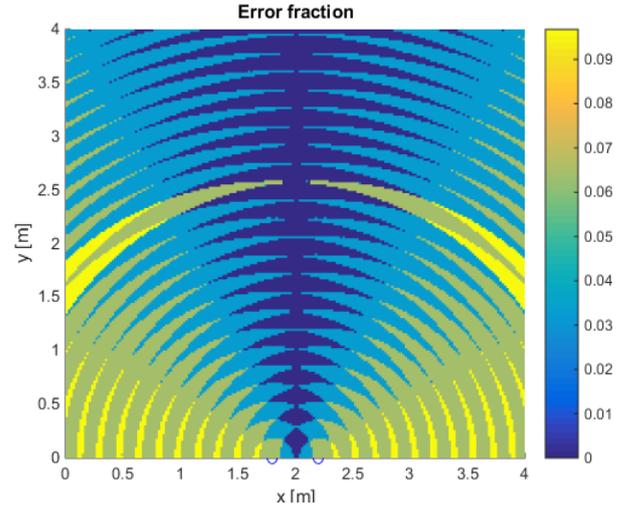

Fig. 3 – Fraction of TDoA measurements affected by ambiguities, before applying the correction heuristic, for a 2-anchor system.

Moreover, the anchors, being fixed and potentially close to each other, may distribute a unified trigger using a wired connection, as shown in the diagram of Fig. 1(b). In this case, in a practical microcontroller-based implementation, the trigger signal for starting the sampling procedure could be given by the rising edge of a logic signal. A polling procedure might be used at the receiver for assessing the status of the trigger signal. Due to this polling procedure, there may be a time delay between the rising edge of the trigger signal and the actual sampling start, which may be different for every receiver. However, such delay would be of the order of microseconds for microcontrollers with clocks in the megahertz range. Therefore, this would cause submillimeter-order deviations in the estimated range, resulting in a negligible contribution to the positioning error.

### C. A heuristic for identifying and correcting the ambiguity

To address the ambiguity issue described in the previous subsection, a simple heuristic method is proposed, which is based on performing elementary operations on the estimated range difference between the mobile node and two anchors. The fundamental goal is to correct TDoA measurements when it is detected that they are not related to the same pulse. Such detection is performed based on constraints given by the known distance between the anchors. The detailed procedure is illustrated in Algorithm 1, where a distance margin is introduced to improve robustness against noise and to reduce false positives. Note that, for dealing with positioning systems consisting of more than two anchors, the algorithm is applied to all anchor couples between which the TDoA is measured.

**Algorithm 1**: *Estimate range difference between the mobile node and two anchors in the presence of ambiguities*

1. **Assume** known distance between anchors $D$, propagation speed $v$, pulse repetition interval $T_C$
2. **Select** distance margin $d_M$
3. **Get** initial range difference estimate $d$
4. **if** $|d| > D + d_M$ **then**
5.    **if** $|d| > 0$ **then**
6.       $d \leftarrow d - T_C \cdot v$
7.    **else**
8.       $d \leftarrow d + T_C \cdot v$
9.    **end if**
10. **end if**
11. **return** $d$

## III. NUMERICAL SIMULATION RESULTS

To investigate the applicability of the proposed method in practical TDoA ultrasound positioning scenarios, numerical simulations are conducted. Positioning in two dimensions using TDoA multilateration on the same gridded area as in Fig. 3 is performed. The transmitted signal is a unit-amplitude linear chirp of with frequency spanning the [38 42] kHz interval, cyclically repeating itself with a period of 15 ms. The position of the mobile transmitter (TX) node is varied along the grid, while three receiver (RX) nodes (i.e. the anchors) are placed at fixed positions. The ideal propagation delay between the transmitter and the receiver is computed using the speed of sound, set at a value of $v = 343$ m/s.

To simulate the signal reception by the RX nodes, the chirp signal is delayed by this ideal propagation delay, and by the latency between the master trigger and the start of the observation window. The path loss model described in [12] is employed to simulate the attenuation of the signal in the transmitter-receiver chain. The numerical values of the parameters of such model are set based on the specifications of the commonly used MA40S4R sensor [22]. Additionally, the effect of sensor directivity is considered by means of an attenuation factor that is dependent on the angle between the direction of maximum directivity of RX and the line joining TX and RX. Such attenuation is calculated according to the directivity specifications in [22]. In the simulation, the sensors are assumed to be oriented so that their direction of maximum directivity is parallel to the y axis. Furthermore, white Gaussian noise, with zero mean and standard deviation σ, is added to the delayed signals at the receivers. Two different noise levels are considered: σ = 0.01, corresponding to an SNR of approximately -4 dB at the maximum distance in the considered grid (about 4.5 m), and σ = 0.001, corresponding to a minimum SNR of 16 dB.

Each RX node measures the time of arrival, as in [12], by applying a processing method that is based on cross-correlation between the received signal and a template of the transmitted signal. Then, the differences between the time-of-arrival measurement results are computed, to effectively eliminate the impact of the unknown latency. Subsequently, TDoA positioning is performed, according to the closed-form algorithm described in [23], for the special case of a linear array of three sensors.

Results are shown in Fig. 4 – 7, where the error fraction is plotted at each point in the grid. Such fraction is calculated as the number of error events divided by the number of trials at each point. As described in Section II-B, each trial is characterized by a different latency between the start of transmission and the start of observation, and there is a total of 31 trials at each grid point. An error event is defined as a distance greater than 1 cm between estimated and true position.

By comparing Fig. 4-5 with Fig. 6-7, it can be noticed that the application of the heuristics is beneficial. In particular, it allows for reducing the error fraction in the central portion of the considered area, achieving a large portion of error-free performance with centimeter-level accuracy. The performance improvement is less perceivable in areas to the left and to the right of the beacons' set, because when the mobile node is close to co-linearity with the beacons, the TDoA problem is ill-conditioned. Conversely, when the mobile node is in front of the beacons, a maximum operational range can be identified, due to signal attenuation and noise level. The performance of the proposed heuristics is summarized in the empirical cumulative distribution function (CDF) plots of Fig. 8, where it is shown that the application of the correction heuristics improves performance in terms of positioning error, in both the σ = 0.01 and σ = 0.001 cases. Therefore, the proposed heuristic is proven to be effective even in presence of noise.

## IV. CONCLUSIONS

A system architecture was analyzed, based on wireless nodes capable of consumer grade RF connectivity, performing ultrasound TDoF measurements. Performance issues caused by both noise and asynchronous operations were investigated, and a methodology to reduce the impact of asynchronous operation was presented, assessing the performance improvement by means of numerical simulations.

### ACKNOWLEDGMENT

This research activity was funded through grant PRIN 2015C37B25 by the Italian Ministry of Instruction, University and Research (MIUR), whose support the authors gratefully acknowledge.

### REFERENCES

[1] B. Wu and Q. Huang, "A Kinect-based automatic ultrasound scanning system," *2016 International Conference on Advanced Robotics and Mechatronics (ICARM)*, Macau, China, 2016, pp. 585-590. doi: 10.1109/ICARM.2016.7606986

[2] B. Meng and J. Liu, "Robotic ultrasound scanning for deep venous thrombosis detection using RGB-D sensor," *Cyber Technology in Automation, Control, and Intelligent Systems (CYBER), 2015 IEEE International Conference on*, Shenyang, 2015, pp. 482-486. doi: 10.1109/CYBER.2015.7287986.

[3] A. Mancini, E. Frontoni and P. Zingaretti, "Embedded Multisensor System for Safe Point-to-Point Navigation of Impaired Users," in *IEEE Transactions on Intelligent Transportation Systems*, vol. 16, no. 6, pp. 3543-3555, Dec. 2015.doi: 10.1109/TITS.2015.2489261.

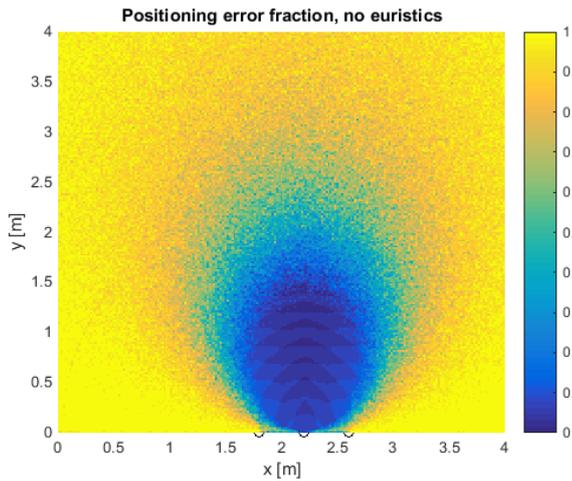

Fig. 4 – Fraction of measurements affected by ambiguities, before applying the correction heuristic, for σ = 0.01.

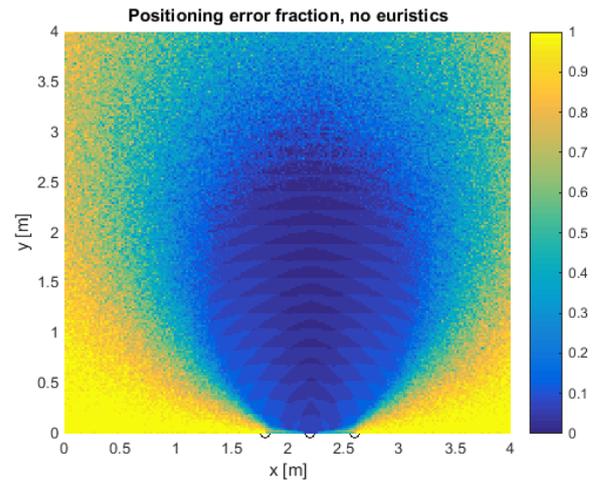

Fig. 5 – Fraction of measurements affected by ambiguities, before applying the correction heuristic, for σ = 0.001.

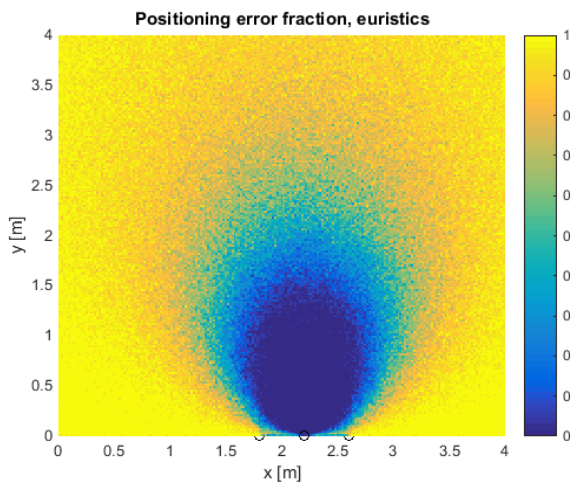

Fig. 6 – Fraction of measurements affected by ambiguities, after applying the correction heuristic, for σ = 0.01.

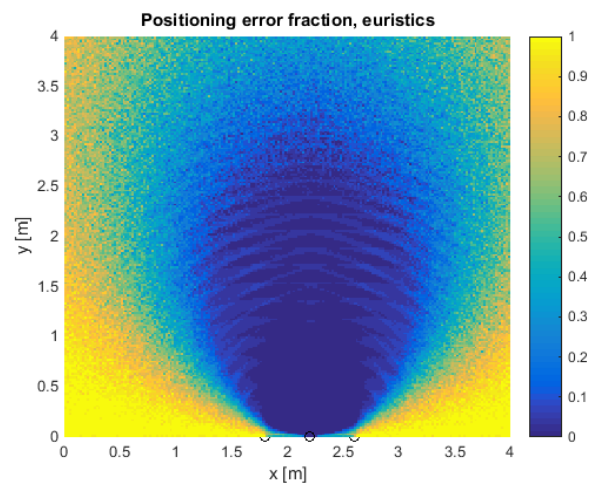

Fig. 7 – Fraction of measurements affected by ambiguities, after applying the correction heuristic, for σ = 0.001.

[4] R. Mautz, "The challenges of indoor environments and specification on some alternative positioning systems," *Proc. of 6th IEEE Workshop on Positioning, Navigation and Communication*, WPNC 2009, pp. 29–36, DOI 10.1109/WPNC.2009.4907800.

[5] H. Yucel, T. Ozkir, R. Edizkan, A. Yazici, "Development of indoor positioning system with ultrasonic and infrared signals," *Proc. of 2012 IEEE International Symposium on Innovations in Intelligent Systems and Applications (INISTA)*, 2012, pp. 1-4, DOI 10.1109/INISTA.2012.6246983.

[6] J.R. Gonzalez and C. J. Bleakely, "High-Precision Robust Broadband Ultrasonic Location and Orientation Estimation," *IEEE Journal on Selected topics in Signal Processing*, Vol. 3, no. 5, October 2009. DOI 10.1109/JSTSP.2009.2027795.

[7] D. Hauschildt, N. Kirchhof, "Improving indoor position estimation by combining active TDOA ultrasound and passive thermal infrared localization," *Proc. IEEE 2011 8th Workshop on Positioning Navigation and Communication (WPNC)*, April 7-8 2011, Dresden, Germany, pp. 94-99, DOI: 10.1109/WPNC.2011.5961022.

[8] N. B. Priyanta, A. Chakraborty, H. Balakrishnan, "The Cricket Location Support System," *Proc. of 6th ACM MOBICOM*, Boston, MA, USA, 2000.

[9] Y. Fukuju, M. Minami, Hiroyuki Morikawa, T. Aoyama, "DOLPHIN: an autonomous indoor positioning system in ubiquitous computing environment," *Proc. IEEE Workshop on Software Technologies for*

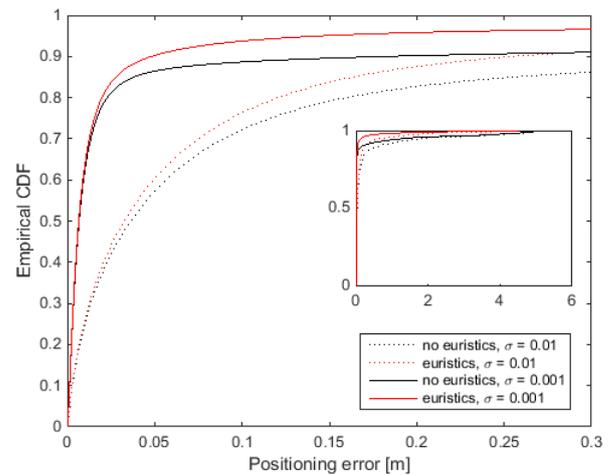

Fig. 8 – Numerical simulation results. Empirical CDF curves of the positioning error for the σ = 0.01 and σ = 0.001 cases.

*Future Embedded Systems*, Japan, 2003, pp. 53-56, DOI 10.1109/WSTFES.2003.1201360.

[10] Kenji Nakahira, Tetsuji Kodama, Takeshi Furukashi, Hiroshi Maeda, "Design of Digital Polarity Correlators in a Multiple-User Sonar Ranging systems," *IEEE Trans. on Instrumentation and Measurement*, Vol. 54, No. 1, February 2005, DOI 10.1109/TIM.2004.839753.

[11] C. Medina, J. C. Segura, A. De la Torre, "Ultrasound Indoor Positioning Based on a Low-Power Wireless Sensor Network Providing Sub-Centimeter Accuracy," *Sensors* 2013, 13, 3501-3526; doi:10.3390/s130303501.

[12] A. De Angelis, A. Moschitta, P. Carbone, M. Calderini, S. Neri, R. Borgna, M. Peppucci, "Design and Characterization of a Portable Ultrasonic Indoor 3-D Positioning System," *IEEE Transactions on Instrumentation and Measurement*, Vol. 64, No. 10, Pp. 2616 – 2625, Oct. 2015. DOI: 10.1109/TIM.2015.2427892.

[13] M. C. Pérez, D. Gualda, J. M. Villadangos, J. Ureña, P. Pajuelo, E. Díaz, E. García, "Android application for indoor positioning of mobile devices using ultrasonic signals," *Proc. International Conference on Indoor Positioning and Indoor Navigation (IPIN)*, Alcala de Henares, 2016, pp. 1-7, doi: 10.1109/IPIN.2016.7743628.

[14] S. Flores, J. Geiß and M. Vossiek, "An ultrasonic sensor network for high-quality range-bearing-based indoor positioning," *Proc. IEEE/ION Position, Location and Navigation Symposium (PLANS)*, Savannah, GA, 2016, pp. 572-576. doi: 10.1109/PLANS.2016.7479747.

[15] D. Macii, A. Ageev and A. Somov, "Power consumption reduction in Wireless Sensor Networks through optimal synchronization," 2009 IEEE Instrumentation and Measurement Technology Conference, Singapore, 2009, pp. 1346-1351.

[16] D. Macii, F. Trenti and P. Pivato, "A robust wireless proximity detection technique based on RSS and ToF measurements," 2011 IEEE International Workshop on Measurements and Networking Proceedings (M&N), Anacapri, 2011, pp. 31-36.

[17] Yi Zhao, J.R. Smith, "A battery-free RFID-based indoor acoustic localization platform,", *Proc. IEEE International Conference on RFID*, pp.110-117, April 30-May 2 2013, doi: 10.1109/RFID.2013.6548143.

[18] Bluetooth 4.0 Core Specification, www.bluetooth.com

[19] A. De Angelis, M. Dionigi, A. Moschitta, P. Carbone, E. Sisinni, P. Ferrari, A. Flammini, S. Rinaldi, "On the Use of Magnetically Coupled Resonators for Chirp-Based Timestamping," in *IEEE Transactions on Instrumentation and Measurement,* vol. 64, no. 12, pp. 3536-3544, Dec. 2015. doi: 10.1109/TIM.2015.2463332.

[20] Y. C. Wu, Q. Chaudhari and E. Serpedin, "Clock Synchronization of Wireless Sensor Networks," in *IEEE Signal Processing Magazine*, vol. 28, no. 1, pp. 124-138, Jan. 2011. doi: 10.1109/MSP.2010.938757.

[21] L. Ferrigno, V. Paciello and A. Pietrosanto, "Experimental Characterization of Synchronization Protocols for Instrument Wireless Interface," in *IEEE Transactions on Instrumentation and Measurement*, vol. 60, no. 3, pp. 1037-1046, March 2011. doi: 10.1109/TIM.2010.2060224.

[22] Murata Manufacturing co., ltd., *MA40S4R Ultrasonic Sensor Application Manual*. 2016.

[23] Y. T. Chan and K. C. Ho, "A simple and efficient estimator for hyperbolic location," in *IEEE Transactions on Signal Processing*, vol. 42, no. 8, pp. 1905-1915, Aug 1994, doi: 10.1109/78.301830.